Title: Vibronic polarons: comments on a model for the colossal field-resistance effects in manganites
Author: Mladen Georgiev (Institute of Solid State Physics, Bulgarian Academy of Sciences 1784 Sofia, Bulgaria)
Comments: 8 pages with 4 figures, all pdf format
Subj-class: physics

In addition to mechanisms already proposed to account for the formation in manganites of a small-polaron superlattice above the Curie temperature $T_C$ and to a metallic-like sea of large polarons below $T_C$, we now consider other observed colossal-resistance inducing fields. Indeed, the order→disorder small-to-large polaron transition is initiated at $T > T_C$ by overcritical external factors (magnetic, electric, photon, or strain fields). We attribute the charge-ordered phase formation to the occurrence of strong dipolar binding of vibronic small polarons, arising from the phonon coupling of highly polarizable two-level orbital systems. The latter species having associated inherent electric and magnetic off-center dipoles largely controlled by optical or mechanical excitation, they couple to the external field thereby leading to the observed colossal effects. The metallic-like phase appears as a result of polaron band widening in an external field which at some stage breaks down the small-polaron condition. We present numerical calculations of the critical field strengths.

1. Foreword

The colossal field-resistance effects (CFR) in manganites pose a challenge to the theoretically minded physicist who often looks at them as a phenomenon to be unraveled rather than as a means to finding new applications. So far, there have been four such effects: magneto-resistance, electro-resistance, photon-resistance, and stress-resistance reported by various experimental groups [1-4]. At this point one may not force too much one's imagination to predict future observations of acousto-resistance, tribo-resistance, etc. phenomena as well. A field-resistance effect occurs as a colossal drop in electric resistance of a manganite when placed in a strong overcritical external field.

Polarons have long been suspected to play a basic role in the phenomenon, though it has not been before recently that more fundamental investigations have been made to decide the issue [5]. It has been found that in a $Mn^{3+}$ magnetic ion manganite (currently widely cited example of $Pr_{1-x}Ca_xMnO_3$), the Curie temperature $T_C$ is the boundary between a metallic-like lower temperature ($T < T_C$) disordered phase and an insulating higher temperature ($T > T_C$) ordered phase. The electron spins arrange over the $Mn^{3+}$-$Mn^{4+}$ manifold ferromagnetically at $T < T_C$ and antiferromagnetically at $T > T_C$ (perhaps paramagnetically too at $T \gg T_C$). The polarons are large below $T_C$ and small above $T_C$. There is a peculiarity in that the small polarons arrange in a charge-ordered state, forming what may be viewed as a small-polaron superlattice, rather than in a vitreous state preserving the long-range order only, or in any other disordered state. The realization that the disordered large-polaron phase precedes the ordered small-polaron phase as T is raised along the temperature axis is a direct challenge to chemical physicists who have long established the rules the other way round [6]. The external field of overcritical strength applied to the manganite at $T > T_C$ incites a phase transition of the order→disorder type which changes the resistance drastically from the high insulating state to the low metallic state value [5].

We have lately suggested that vibronic off-center polarons may eventually prove the species appropriate for inciting the observed field-resistance effects. As a matter of fact, $Mn^{3+}$ is not only magnetic but is a Jahn-Teller (JT) ion as well. Namely, vibronic JT $E_g$-mode mixing causes the splitting of $Mn^{3+}$ $e_g$ orbital levels (singly occupied) in the parent $PrMnO_3$ phase causing in- plane elongation of the $MnO_6$ octahedron [7]. In order to form, vibronic Pseudo-Jahn-Teller (PJT) polarons require the availability of two different-parity orbital electronic states (one of which, say $t_{1u}$, singly occupied) to be mixed by an odd-parity vibrational mode (say $T_{1u}$) [8]. Indeed large and small PJT polarons would appear consequentially as the electron–to-mode coupling strength is increased to bring about the vibronic mixing of orbital electronic states in the two-level system.

It is the vibronic PJT polarons that may do the job of coupling to the external electric and magnetic fields, since the individual species are both electrostatically polarizable [8] and carry magnetic dipole moments [9] arising from their rotation-like reorientation over the off-center sites. To account for the colossal stress-resistance effect, we note that a uniform stress field would increase the polaron bandgap, thereby turning the balance towards the large polarons. Now, the insulating state would be destroyed once most of the small polarons have converted to large ones. The light-induced colossal resistance effect is somewhat more complicated. Our former analyses have indicated that the reorientation of off-center polarons is controlled by a dual branch potential energy surface corresponding to q < 0 (lower branch) and q > 0 (higher branch), respectively, where q is Mathieu's parameter [10]. The eigenenergy bands $(a_0,a_1)$, $(b_1,b_2)$, $(a_2,a_3)$, $(b_3,b_4)$, … (q<0) and $(a_1,b_1)$, $(a_2,b_2)$, $(a_3,b_3)$, $(a_4,b_4)$, … (q>0) are composed of definite parity states (lower branch) and of mixed parity states (upper branch), respectively. Now, we have shown that reorientation is only possible within the definite-parity manifold, while it is frozen-in in the mixed parity manifold. It follows that once lifted to an upper branch energy band, principally via optical excitation, the reorientating entity will become frozen in over there. An obvious consequence is the immobilization of an off-center orbital current following the excitation. This would destroy a magnetic dipole and the magnetic part of the binding energy and may cause the demolition of the small-polaron superlattice.

In any event, a photon induced colossal resistance effect has been witnessed beautifully via the appearance of metallic reflection along the illuminated path on the manganite surface [3].

The electronic energy gap associated with the two-level system is exceedingly close to four times the coupling energy for large polarons ($E_{gap} \geq 4E_{JT}$), while for small polarons the limitation is more stringent ($E_{gap} \gg 4E_{JT}$) [8]. The useful quantity to distinguish between large and small polarons is the polaron binding energy which relates the terms of "large" and "small" species to their sizes: $E_{bind} = E_{JT}[1 + (E_{gap}/4E_{JT})^2]$. For large polarons this gives $E_{bind} \sim ½E_{gap}$, while for small polarons it yields $E_{bind} \sim E_{JT}$ ($E_{JT} \gg ¼E_{gap}$). Iit may not be hard to suggest the opposite parity electronic orbital states associated with the proposed vibronic polarons. By analogy with high-$T_c$ layered perovskites, PJT mixing may occur of $2p_z$ and $3d_z^2$ orbitals along the O(1)-Mn(0)-O(2) tri-atomic segment. The vibronic phenomena relating to the configurational symmetry of manganite clusters, another aspect is the effect of magnetic ordering on the conductivity of a cluster. Concomitantly we assume that wide bandwidth ferromagnetic arrangements are inherent of high conductivity (metallic) large polarons, while narrow bands of low conductivity antiferromagnetic arrangements are typical for the small polarons.

## 2. Binding energy of a vibronic small-polaron superlattice

The binding energy of the vibronic small-polaron superlattice may be envisaged composed of an enhanced Van der Waals electrostatic part [11]

$$V_{VdW} \sim \tfrac{1}{2}\Delta E(F)[\alpha(F)/\kappa]^2 \Sigma_{ij}(1/R_{ij}^6) = \Delta E(F)[p_{12}^2/3\Delta E(F)/\kappa]^2 \alpha_{VdW}/R^6 \qquad (1)$$

where $\alpha_{VdW} = \Sigma_{ij}(R/R_{ij})^6$ is the VdW lattice-sum constant, $\alpha(F) = p_{12}^2/3\Delta E(F)$ is the field-dependent polarizability, $\Delta E(F)$ is the field-dependent two-level tunneling splitting, $\kappa$ is a dielectric constant, R is the small-polaron lattice constant. Incorporating the magnetic-dipole part

$$V_{MD} \sim [3(\boldsymbol{\mu}_i.\mathbf{R}_{ij0})(\boldsymbol{\mu}_j.\mathbf{R}_{ij0}) - (\boldsymbol{\mu}_i.\boldsymbol{\mu}_j)]\alpha_{MD}/\mu_0 R^3 \qquad (2)$$

where $\boldsymbol{\mu}_i$ are the magnetic dipoles, $\mathbf{R}_{ij0}$ is the unit vector along the dipolar separation $R_{ij}$, $\mu_0$ is the permeability, $\alpha_{MD} = \Sigma_{ij}(R/R_{ij})^3$ is the magnetic lattice-sum constant. Summing up we get for the binding energy of a vibronic small-polaron superlattice:

$$V_{bind} = \{\tfrac{1}{2}\Delta E(F)[\alpha(F)/\kappa]^2 \alpha_{VdW}/R^3 + [3(\boldsymbol{\mu}_i.\mathbf{R}_{ij0})(\boldsymbol{\mu}_j.\mathbf{R}_{ij0}) - (\boldsymbol{\mu}_i.\boldsymbol{\mu}_j)](\alpha_{MD}/\mu_0)\}(1/R^3)$$

$$= \{\tfrac{1}{2}[(p_{12}^4/3\kappa^2)/\Delta E(F)](\alpha_{VdW}/R^3) + [3(\boldsymbol{\mu}_i.\mathbf{R}_{ij0})(\boldsymbol{\mu}_j.\mathbf{R}_{ij0}) - (\boldsymbol{\mu}_i.\boldsymbol{\mu}_j)](\alpha_{MD}/\mu_0)\}(1/R^3) \qquad (3)$$

The critical field strength will be obtained from $V_{bind} \leq k_B T$ which gives

$$\{\tfrac{1}{2}[(p_{12}^4/3\kappa^2)/\Delta E(F)](\alpha_{VdW}/R^3) + [3(\boldsymbol{\mu}_i.\mathbf{R}_{ij0})(\boldsymbol{\mu}_j.\mathbf{R}_{ij0}) - (\boldsymbol{\mu}_i.\boldsymbol{\mu}_j)](\alpha_{MD}/\mu_0)\}(1/R^3) \leq k_B T \qquad (4)$$

whereby we get

$$\Delta E(F) \geq (p_{12}^4/3\kappa^2)(\alpha_{VdW}/R^6)/\{k_B T - [3(\boldsymbol{\mu}_i.\mathbf{R}_{ij0})(\boldsymbol{\mu}_j.\mathbf{R}_{ij0}) - (\boldsymbol{\mu}_i.\boldsymbol{\mu}_j)](\alpha_{MD}/\mu_0)(1/R^3)\} \qquad (5)$$

The field-dependent tunneling splitting is

$$\Delta E(F) = \sqrt{[\Delta E(0)^2 + H_F^2]} \qquad (6)$$

where $\Delta E(0)$ is the splitting in the absence of an external field:

$$\Delta E(0) = E_{gap} \exp(-2E_{JT}/\eta\omega) \qquad (7)$$

(Holstein's squeezed small polaron gap), where $\omega$ is the coupled vibrational mode frequency, $\eta = h/2\pi$, while the field coupling terms are:

$H_F = -\mathbf{p}_{12}.\mathbf{F}$ (electric), $H_F = -\boldsymbol{\mu}_i.\mathbf{H}$ (magnetic), $H_F = -\boldsymbol{\varepsilon}_i.\mathbf{S}$ (stress),

$$H_F = -\mathbf{p}_E.\mathbf{E} \text{ (light)}(\mu_k = 0) \qquad (8)$$

For any particular situation, the critical field strength $\mathbf{F}_C$ is to be calculated by equations (5) through (8). For an electric-field induced colossal effect, we obtain

$$F_C = \{[(p_{12}^4/3\kappa^2)(\alpha_{VdW}/R^6)/\{k_B T - [3(\boldsymbol{\mu}_i \cdot \mathbf{R}_{ij0})(\boldsymbol{\mu}_j \cdot \mathbf{R}_{ij0}) - (\boldsymbol{\mu}_i \cdot \boldsymbol{\mu}_j)](\alpha_{MD}/\mu_0)(1/R^3)\}]^2 - \Delta E(0)^2\}^{1/2}$$

$/ p_{12F}$ (9)

where $p_{12F}$ is the projection along the field direction of the electrostatic mixing dipole. The critical magnetic, strain, and light fields will be derived along similar lines, though in the general denominator of equation (9) $p_{12F}$ will have to be substituted for by $\mu_{iH}$, $\varepsilon_{iS}$, $p_{EE}$, respectively.

Clearly, the sole formation of a small-polaron superlattice over Mn sites, e.g. in $Pr_{1-x}Ca_xMnO_3$ as a typical case, does not suffice, for there may be other aspects of the general conditions for the occurrence of a colossal resistance effect not covered by the electrostatic and magneto-static orderings alone. For instance, if part of the $Mn^{3+}$ ions are substituted for by other magnetic ions, such as $Fe^{3+}$ of nearly the same ionic radii, to form $Bi_{0.5}Sr_{0.5}Fe_xMn_{1-x}O_3$, then the colossal effect is not observed for x > 0.4. (The superstructure is orthorhombic for $x \leq 0.3$ and cubic for $x > 0.4$ [12]). Apparently the current field theory is missing something which may prove important for understanding the phenomenon in $Bi_{0.5}Sr_{0.5}Fe_xMn_{1-x}O_3$ ($0 \leq x \leq 1.0$).

### 3. Colossal field-resistance effect

The field-resistance effects are shown in Figure 1 for a strong magnetic field. We see both the insulating range to the right and the metallic range to the left of the peak temperature $T_C$.

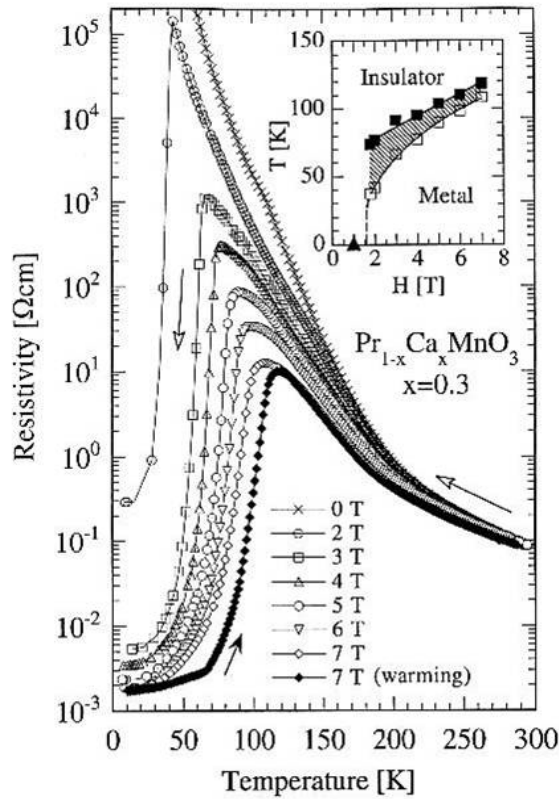

Figure 1. A series of experimental resistivity vs. temperature curves at various values of the external magnetic field between 0÷7 T. Internet data by Y. Tokura at the University of Tokyo

We stress time and again our opinion that Holstein's polarons, which couple to symmetric modes, may react to the external magnetic field by virtue of their off-center orbital magnetic moments. Holstein's small polarons have been considered in the literature within the context of the colossal resistivity problem [13,14]. An issue arises when considering the eventual manifestation of a parallel colossal electro-resistivity effect, that is, the observation of similar changes of resistance in strong electric fields. (So far, only single measurements though no systematic studies similar to Figure 1 have been reported [5].) If confirmed, the electric field effects will face difficulties to be attributed to Holstein's small polarons, since they do not carry electric dipole moments to couple to the external field. Nevertheless, off-center vibronic polarons appear to be the more appropriate species, since they carry both off-center electric and magnetic dipoles. The electric-field induced colossal resistance effect, as predicted by our off-center vibronic-polaron model, is shown in Figure 2 where we have plotted the critical field strength $E_C$ by equation (9), physically, at $E_C$ the binding energy falls down lower than $k_BT$. We have commented above on the possibility to incorporate stress- and light- induced colossal effects (so far single observations too), since they both will tend to reduce the small-polaron binding energy.

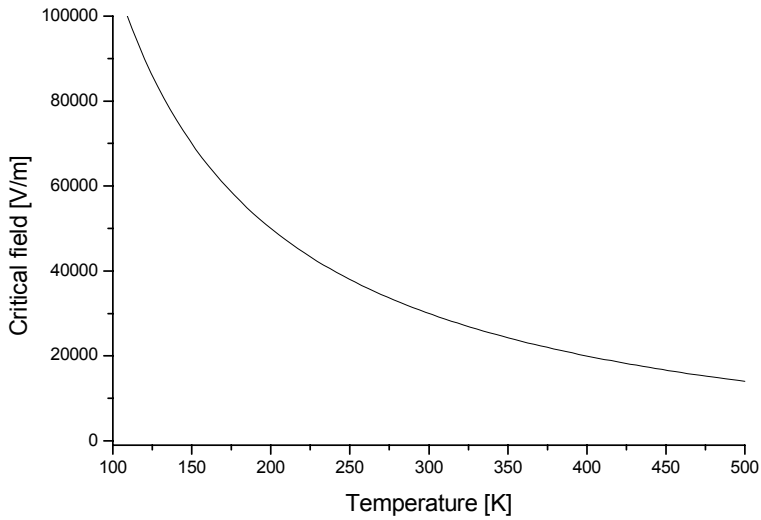

Figure 2. Temperature dependence of the critical field $F_C = \{[(p_{12}^4/3\kappa^2)(\alpha_{VdW}/R^6) / k_BT]^2 - \Delta E(0)^2\}^{1/2} / p_{12F}$ for the colossal electro-resistance effect in a manganite perovskite material with basic-lattice cubic parameter $a$ = 4 Å, superlattice cubic parameter R = $3a$ = 12 Å, mixing electric dipole $p_{12F}$ = 1 eÅ, lattice sum constant $\alpha_{VdW}$ = 6, dielectric constant $\kappa$ = 5. The magnetic effects on the binding and coupling energies have been disregarded for simplicity setting the off-center magnetic dipole to $\mu$ = 0. From Reference [11].

The magnetic dipoles have been evaluated using Mathieu's functions, the eigenstates of the reorientating off-center oscillators [9]:

$$\mu = (4\pi\mu_0/c)(\pi\rho^2)\Omega_{rot} = 1.32\times10^{-15} \mu_0\rho^2 \Omega_{rot} \quad [\text{Å}^2\text{THz}] \qquad (10)$$

where $\rho$ is the off-center radius and $\Omega_{rot}$ is the rotational frequency in the particular rotational band, $\mu_0$ is the medium permeability. To disregard $\mu$ would mean to discard the off-center displacements, while to do so with $\Omega_{rot}$ would mean discarding the reorientation over the off-center sites. None of these is permissible for off-center vibronic polarons.

4. Further comments

The electronic states involved in the Jahn-Teller effects have been identified as the $Mn^{3+}$ $e_g$ states mixed by the $E_g$-mode $Q_\varepsilon$ and $Q_\theta$ coordinates [14]. The ($\varepsilon$-$\theta$) mixing leads to a JT splitting of the $e_g$ doublet ($3d_x^2$, $3d_y^2$). By analogy with the high-$T_c$ superconductor, we attribute the expected PJT coupling to the mixing of the $3d_z^2$ orbital state with the linear combination of the oxygen frame along the z-axis, namely $2p_z(1)$&$2p_z(2)$ ($t_{1uz}(1)$&$t_{1uz}(2)$) by the $T_{1uz}$ vibrational mode [15]. Figure 3 shows schematically the spatial dependencies of the electron wavefunction across the $2p_z(1)$-$3d_z^2(0)$-$2p_z(2)$ triatomic segment in the $La_{2-x}Sr_xCuO_4$ high-Tc superconductor. It suggests just how the odd-parity vibration modulates the z-(c-) axis link to transfer electron density along the z-axis in $Pr_{1-x}Ca_xMnO_3$.

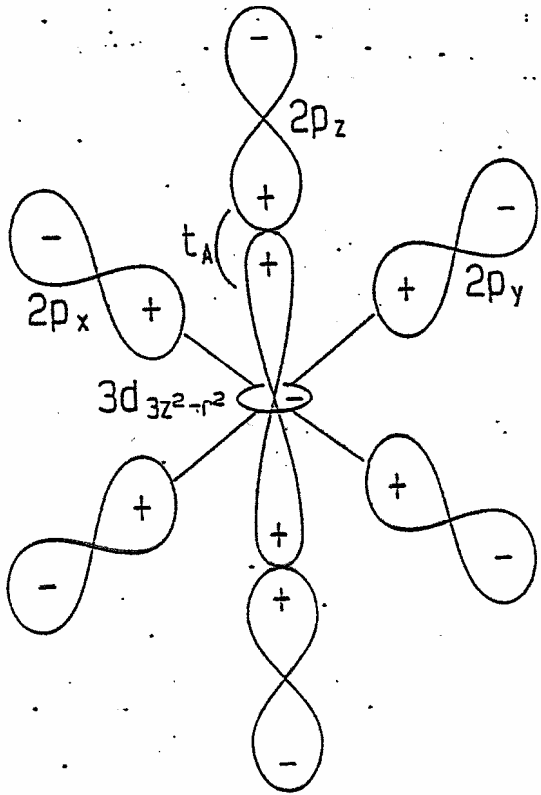

Figure 3. Ionic orbitals used in the PJT vibronic model for $La_{2-x}Sr_xCuO_4$: the in-plane $a_{1g}$ O(P) ligand, the $3d_{3z^2-r^2}$ $Cu^{2+}$(P), and the two apex $2p_z$ O(A). $t_A$ is the hopping between Cu(P) and O(A). O(A) vibrations: even O(A)→–Cu–←O(A) and odd O(A)→–Cu—O(A)→. Ref. [16].

The allowed energy-band structure of the reorientating off-center ion ($Mn^{4+}$) ($Mn^{3+}$ + hole polaron) consists of bands at q < 0 coupled to the (two-site) adiabatic ground state, as well as ones at q > 0 coupled to the (two-site) adiabatic excited state. The q-dependences of the allowed bands are shown in Figure 4. The former are composed of definite parity states: even, odd, etc., as mentioned above, the latter are mixed parity. An analysis based on the occurrence probability $W_{conf}(E_n)$ for the transition from one reorientational site to the neighboring site indicates that the intraband transition is only nonvanishing when within a definite-parity band. At the same time, the intraband probability does vanish when within a mixed parity band [10]. It follows that reorientation will only occur when the particle is in the adiabatic ground state, while the particle will stay frozen if in the excited adiabatic state. Accordingly, only the $q < 0$ bands are genuinely rotational, while the $q > 0$ bands are not.

We do not invalidate the possibility that, on optical excitation, the off-center manganese has been lifted to a mixed-parity state where it has remained frozen. (In addition, the entity may be transferred to the adiabatic ground state by a subsequent nonradiative process [17].) Such optical transitions are not ruled out. For the time being, we do not have much prior knowledge of the way the light-induced colossal-resistance experiment has been done.

As regards the charge ordering in the CFR-active manganite, doping with divalent impurities, such as $Ca^{2+}$, should be accompanied by the occurrence of an equivalent number of charge compensating species. These are holes within the temperature range of present interest, or oxygen vacancies at higher temperatures. The holes may give rise to a random phase yielding electric conductivity to the manganite. The vacancies may be too heavy to move effectively. Eventually, the positively charged oxygen vacancies may join the manifold of negatively charged divalent cations and the hole small polarons to form kind of a lattice structure. If the latter is to bear the features of the mother crystal, the latter structure might be a cubic or orthorhombic superlattice composed of divalent ions and hole polarons, and a small amount of compensating oxygen vacancies, perhaps too small to play any essential role near the Curie temperature. We believe the presence of divalent ions coherent with the polaron ordering should be accounted for; otherwise the superstructure should lack stability. For this reason, we refer to the superstructure as superlattice rather than as merely charge ordering..

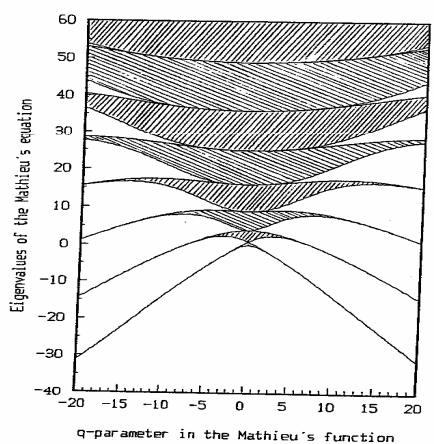

Figure 4. The q-dependence of the reorientational energy bands at the off-center ion should it sit at a Mn site in manganites. Only the bands at q < 0 are genuinely rotational, while those at q > 0 are allowed though not rotational. Only transitions to and fro definite parity bands at q < 0 are permissible if they should concern rotating entities. (Courtesy of Dr. D. Batovsky, Ascension University, Bangkok.).

In the above respect, it would be very interesting to learn if the colossal resistance effects could be produced in photodoped manganites, that is, ones in which free charge carriers were generated by photoconductivity rather than by chemical doping. Eventually, photoelectrons and photoholes will be phase-segregated as they are in photodoped high-$T_c$ superconductors.